\documentstyle[12pt]{article}
\newcommand{\bee}{\begin{equation}}
\newcommand{\ee}{\end{equation}}

\newcommand{\R}{\rm I\kern-.2emR}
\newcommand{\C}{\rm \kern.25em\vrule height1.4ex
 depth-.12ex width.06em\kern-.31em C}
\newcommand{\N}{{\rm I\kern-.16em N}}
\newcommand{\Z}{{\rm Z\kern-.35em Z}}
\begin{document}
\thispagestyle{empty}
\begin{flushright}
July 1994
\end{flushright}
\bigskip\bigskip\begin{center}
Reply to M.Campostrinis's and P.Rossi's Comment on our paper
\vskip5mm
{\bf \Large{`Nonuniformity of the $1/N$ Expansion for
Two-Dimensional $O(N)$ Models'}}
\vskip 1.0truecm
{\bf Adrian Patrascioiu}
\centerline{Physics Department, University
of Arizona, Tucson AZ 85721, U.S.A.}

and

{\bf Erhard Seiler}
\centerline{Max-Planck-Institut f\"{u}r
 Physik, Werner-Heisenberg-Institut}
\centerline{F\"ohringer Ring 6, 80805 Munich, Germany}
\end{center}
\vskip5mm
The Comment to our paper by Campostrini and Rossi \cite{Campo} calls for
a reply. In our paper we tried to emphasize two points regarding the $1/N$
expansion:

1) Mathematically rigorous results obtained so far can establish {\it only}
that it produces the correct asymptotic expansion at fixed $\tilde\beta$; in
particular they do not guarantee that the expansion is asymptotic to
the true expectation values uniformly in $\tilde\beta$ so that information
about the true behavior for large $\tilde\beta$ can be obtained from it.

2) In $2D$, but not in $1D$, the coefficient of the $1/N$ term in the
expansion of $G(\xi_\infty/2)/G(\xi_\infty/4)$ grows linearly with
$\tilde\beta$. The quantity we consider is clearly `adimensional', as
they call it, but the fact that the coefficient of $1/N$ is unbounded
implies that the expansion is non-uniform, contrary to their statement
No.3.

Campostrini and Rossi have nothing to say about point (1), whose significance
they seem to have missed completely. Indeed in the first part of their
Comment, they discuss their results for the {\it continuum} $O(N)$ model.
At the present time, no such model has been constructed.
Consequently how could one ask
whether the $1/N$ expansion produces the correct asymptotic expansion or not
(see Sect.3 in our paper for the correct definition of asymptoticity),
when the quantities with which the expansion is to be compared have not even
been defined?

They acknowledge that nonuniformity does occur
in $\chi$ and $\xi$ and call it a `predicted
property'. Presumably they mean that the asymptotic scaling formula itself
has this property. But this fact certainly does not make the nonuniformity
any less real. Similarly, they say that they can compute those nonuniformities
analytically; again, that does not detract from the fact that they are there.

Finally, they claim that the nonuniformities in certain quantities
disappear if one expands in $1/(N-2)$ instead of $1/N$. It is easy to see
that this is mathematically impossible and that one can prove the following

\noindent
{\bf Theorem:} {\it Let $F(N,\beta)$ be a function that has an aymptotic
expansion in powers of $1/(N-2)$ that is uniformly asymptotic in $\beta$.
Then $F$ has also an asymptotic expansion in powers of $1/N$ that is uniformly
asymptotic.}

Campostrini and Rossi probably got mislead by the fact that by expanding
in $1/(N-2)$ they can make certain manifestly nonuniform terms vanish.
But uniform asymptoticity is a not a statement about some terms in the
expansion, but about the truncation error.

The failure of Campostrini and Rossi to understand the issue is further
illustrated by their statement: `Adimensional ratios of physical quantities,
computed in the lattice model, show uniform asymptoticity (actually complete
independence of $\beta$ in the scaling limit'. Again, as we stressed in our
paper, to establish the correctness of the expansion one must bound the
truncation error; the finiteness of the expansion coefficients for
$\tilde\beta\to\infty$ is necessary, but not sufficient to establish uniform
asymptoticity.

Campostrini and Rossi ignore completely  point (2) above, which provides
a concrete illustration that in actual fact, in $2D$, the $1/N$ expansion is
non-uniform, at least for certain observables. They are also
misinterpreting the numerical evidence, which shows systematically that
the correlation length and the magnetic susceptibility grow faster than
expected from asymptotic freedom (see Apostolakis et al \cite{Apo}).

In conclusion, the mathematically well defined question is whether in the
lattice $O(N)$ model the $1/N$ expansion remains uniformly asymptotic
for $\tilde\beta\to\infty$?
None of the observations made by Campostrini and Rossi in their Comment
prove that point. To show that there is indeed a problem,
we considered the ratio $G(\xi_\infty/2)/G(\xi_\infty/4)$, which for
$\tilde\beta\to\infty$ approaches a finite value in the spherical model;
we found that the coefficient of the $1/N$ correction to this ratio increases
linearly with $\tilde\beta$ in $2D$, while it remains finite in $1D$. This
fact casts doubt upon prior attempts to prove that $2D$ $O(N)$, $N>2$ models
have a mass gap for any $\beta<\infty$ via the $1/N$ expansion.

\end{document}